\documentclass[12pt]{article}%
\usepackage{ amsmath, graphics, rotating}%

\begin{document}
\large
\begin{center}{\large\bf QUANTUM THEORY AND GALOIS FIELDS}
\end{center}
\vskip 1em \begin{center} {\large Felix M. Lev} \end{center}
\vskip 1em \begin{center} {\it Artwork Conversion Software Inc.,
1201 Morningside Drive, Manhattan Beach, CA 90266, USA 
(E-mail:  felixlev@hotmail.com)} \end{center}
\vskip 1em

{\it Abstract:}
\vskip 0.5em

We discuss the motivation and main results of a
quantum theory over a Galois field (GFQT). The goal
of the paper is to describe main ideas of GFQT in a 
simplest possible way and to give clear
and simple arguments that GFQT is a more natural quantum theory
than the standard one. The paper has been prepared as
a presentation to the ICSSUR' 2005 conference (Besancon, France,
May 2-6, 2005).

\section{Modern physics and spacetime}
\label{S1}

The phenomenon of local quantum field theory (LQFT) has no
analogs in the history of science. There is no branch of
science where so impressive agreements between theory and
experiment have been achieved. The theory has successfully
predicted the existence of many new particles
and even new interactions. It is hard to believe that all 
these achievements
are only coincidences.

At the same time, the level of
mathematical rigor in LQFT is very poor and, as a
result, LQFT has several well known difficulties and
inconsistencies. Probably the main inconsistency is that
local quantum fields are meaningful only as operator-valued
distributions, and products of such objects at the same
spacetime points is ambiguous. The problem of how such
products should be correctly defined is discussed in a wide
literature but a universal solution has not been found yet. 

The absolute majority of physicists believes that agreement 
with experiment is much more important than the lack of 
mathematical rigor. The philosophy of most optimistic
physicists is roughly as follows. The standard model describes 
almost everything and to construct the ultimate theory,
this model should be only modified somehow at Planck
distances. The content of the present paper will probably
be of no interest for such physicists.

At the same time, some famous physicists are not so optimistic.
For example, Weinberg believes \cite{Wein2} that the new theory
may be 'centuries away'. Although he has contributed much
to LQFT and believes that it can be treated 
{\it the way it is}, he also believes 
that it is a {\it low energy approximation to a deeper theory 
that may not even be a field theory,
but something different like a string theory} \cite{Wein}.

Dirac was probably the least optimistic famous physicist.
In his opinion \cite{DirMath}: {\it The agreement
with observation is presumably by coincidence, just like the
original calculation of the hydrogen spectrum with Bohr orbits.
Such coincidences are no reason for turning a blind eye to the
faults of the theory. Quantum electrodynamics is rather like
Klein-Gordon equation. It was built up from physical ideas
that were not correctly incorporated into the theory and it
has no sound mathematical foundation}. 

The main problem is the choice of strategy for constructing a 
new quantum theory. Since nobody knows for sure which strategy 
is the best one, different approaches should be investigated. 
In the present paper we are trying to follow Dirac's
advice given in Ref. \cite{DirMath}: {\it I learned
to distrust all physical concepts as a basis for a theory. 
Instead one should put one's trust in a mathematical scheme, 
even if the scheme does not appear at first sight to be
connected with physics. One should concentrate on
getting an interesting mathematics}.

We believe that quantum theory over a Galois field (GFQT)
is not only an interesting mathematical theory but it
may be a basis for the ultimate quantum physics. The goal 
of the present paper is to describe main
ideas of GFQT in a simplest possible way and to give clear
and simple arguments which hopefully might convince some
physicists that GFQT is a more natural quantum theory
than the standard one. However, before discussing GFQT
we would like to note the following. The physicists
claiming that only agreement with experiment is of any
importance, typically do not pay attention not only
to the lack of rigor in LQFT but also to the fact that
the present approaches to elementary particle theories
are not quite consistent. Let us try to elaborate the
last statement.

If you ask modern physicists whether they believe in
quantum theory, the absolute majority of them will
answer 'yes' without any doubt. Then you could ask
the next question: do you agree with one of the main statements
of quantum theory that every physical quantity should
be related to some (selfadjoint) operator? And again,
the absolute majority will answer 'yes'. But in that
case is time a well defined physical quantity? It has
been known for many years (see e.g. Ref. \cite{time})
that there is no good operator, which can be related to
time. In particular, we cannot construct a state which
is the eigenvector of the time operator with the eigenvalue
-5000 years BC or 2006 years AD.

It is also
well known that, when quantum mechanics is combined with
relativity, there is no operator satisfying all the
properties of the spatial position operator
(see e.g. Ref. \cite{NW}). In other words, the coordinate 
cannot be exactly measured by itself even in situations when
exact measurement is allowed by uncertainty principle.

In the introductory section of the well-known textbook 
\cite{BLP} simple arguments are given that for a particle 
with the mass $m$ the coordinate cannot be measured with 
the accuracy better than the Compton wave length $\hbar/mc$.
Therefore exact measurement is possible only either in the
nonrelativistic limit (when $c\rightarrow \infty$) or
classical limit (when $\hbar\rightarrow 0$). 

In particular, the quantity $x$ in the Lagrangian density 
$L(x)$ is only a parameter which becomes the coordinate in 
the nonrelativistic or classical limit. Note that even
in the standard formulation of LQFT, the Lagrangian is
only an auxiliary tool for constructing Hilbert spaces
and operators. After this construction has been done,
one can safely forget about Lagrangian and concentrate 
his or her efforts on calculating different observables.
As Rosenfeld writes in his memoirs about Bohr 
\cite{Rosenfeld}: {\it His (Bohr) first remark ... was that 
field components taken at definite space-time
points are used in the formalism as idealization 
without immediate physical meaning; the only meaningful
statements of the theory concern averages of such fields 
components over finite space-time regions....Bohr 
certainly never showed any respect
for the noble elegance of a Lagrangian principle.}

The facts that the time and coordinate are not measurable
were known already in 30th of the last
century and became very popular in 60th (recall the
famous Heisenberg S-matrix program). The authors of Ref. 
\cite{BLP} claim that spacetime, Lagrangian and local 
quantum fields are rudimentary notions which will 
disappear in the ultimate quantum theory.
Since that time, no arguments questioning those ideas have
been given, but in view of the great success of gauge
theories in 70th and 80th, these ideas became almost forgotten.

The success of gauge theories and new results in the string
theory have revived the hope that Einstein's dream about
geometrization of physics could be implemented. Einstein said
that the left-hand-side of his equation of 
General Relativity (GR),
containing the Ricci tensor, is made from gold while the
right-hand-side containing the energy-momentum tensor of the
matter is made from wood. Since that time a lot of efforts
have been made to derive physics from geometry of spacetime.
The modern ideas in the (super)string theory are such that 
quantum gravity comes into play at Plank distances and all
the existing interactions can be described if we find 
how the extra dimensions are compactified. These investigations
involve very sophisticated methods of topology, algebraic
geometry etc.

We believe that such investigations might be of mathematical
interests and might give interesting results but cannot
lead to ultimate quantum theory. It is rather obvious that 
geometrical and topological ideas 
originate from our macroscopic experience. For example,
the water in the ocean seems to be continuous and is described
with a good accuracy by equations of hydrodynamics. At the
same time, we understand that this is only an approximation
and in fact the water is discrete. 
As follows from the above discussion, the notion of spacetime
at Planck distances does not have any physical significance
and therefore methods involving geometry, topology, manifolds 
etc. at such distances cannot give a reasonable physics. 

While the notion of spacetime can only be a good approximation
at some conditions, the notion of empty spacetime fully
contradicts the basic principles of quantum theory that
only measurable quantities can have a physical meaning.
Meanwhile the modern theories often begin with the 
background empty spacetime. Many years ago, when quantum
theory was not known, Mach proposed his famous principle, 
according to which the properties of space at a given point 
depend on the distribution of masses in the whole Universe. 
This principle is fully in the spirit of quantum theory.

As described in a wide literature (see e.g. Refs. 
\cite{Mach1,Mach2,Wein2} and references therein), Mach's 
principle was a guiding one for Einstein in developing GR, but
when it has been constructed, it has been realized that GR 
does not contain Mach's principle at all! As noted in Refs. 
\cite{Mach1,Mach2,Wein2}, this problem is not closed.

Consider now how one should define the notion of elementary 
particles. Although particles are observable and fields are not,
in the spirit of LQFT fields are more fundamental
than particles, and a possible definition is as
follows \cite{Wein1}: 'It is simply a particle whose
field appears in the Lagrangian. It does not matter if
it's stable, unstable, heavy, light --- if its field
appears in the Lagrangian then it's elementary,
otherwise it's composite'.

Another approach has been developed by Wigner in his
investigations of unitary irreducible representations
(IRs) of the Poincare group \cite{Wigner}. In view
of this approach, one might postulate that a particle
is elementary if the set of its wave functions is the 
space of a unitary IR of the symmetry group in the given theory.

Although in standard well-known theories (QED, electroweak
theory and QCD) the above approaches are equivalent,
the following problem arises. The symmetry
group is usually chosen as a group of motions of
some classical
manifold. How does this agree with the above discussion
that quantum theory in the operator formulation should
not contain spacetime? A possible answer is as follows.
One can notice that for calculating observables (e.g. the
spectrum of the Hamiltonian) we need in fact not a
representation of the group but a representation of its
Lie algebra by Hermitian operators. After such a
representation has been constructed, we have only
operators acting in the Hilbert space and this is all
we need in the operator approach. The representation
operators of the group are needed only if it is
necessary to calculate some macroscopic
transformation, e.g. time evolution. In the approximation
when classical time is a good approximate parameter,
one can calculate evolution, but nothing guarantees
that this is always the case (for example, it is obviously
unreasonable to describe states of a system every $10^{-1000}sec$ or consider
translations by $10^{-1000}cm$). Let us also note that in
the stationary formulation of scattering theory, the
S-matrix can be defined without any mentioning of
time (see e.g. Ref. \cite{Kato}). For these reasons
we can assume that on quantum level the symmetry
algebra is more fundamental than the symmetry group.

In other words, instead of saying that some operators
satisfy commutation relations of a Lie algebra
$A$ because spacetime $X$ has a group of motions $G$ such
that $A$ is the Lie algebra of $G$, we say that there
exist operators satisfying  commutation
relations of the Lie algebra $A$ such that: for some
operator functions $\{O\}$ of them the classical
limit is a good approximation, a set $X$ of the eigenvalues
of the operators $\{O\}$ represents a classical manifold with
the group of motions $G$ and its Lie algebra is $A$.
This is not of course in the spirit of famous Klein's Erlangen
program or LQFT.

Consider for illustration the well-known example of
nonrelativistic quantum mechanics. Usually
the existence of the Galilei spacetime is assumed from
the beginning. Let $({\bf r},t)$ be the spacetime coordinates
of a particle in that spacetime. Then
the particle momentum operator is $-i\partial/\partial {\bf r}$
and the Hamiltonian describes evolution by the
Schroedinger equation.
In our approach one starts from an IR of the Galilei algebra.
The momentum operator and the Hamiltonian represent four of ten
generators of such a representation. If it is implemented in
a space of functions
$\psi({\bf p})$ then the momentum operator is simply the
operator of multiplication by ${\bf p}$. Then the
position operator can be {\it defined} as
$i\partial/\partial {\bf p}$
and time can be {\it defined} as an evolution
parameter such that
evolution is described by the Schroedinger equation with
the given Hamiltonian. Mathematically the both approaches are
equivalent since they are related to each other by the Fourier
transform. However, the philosophies behind them are
essentially different. In the second approach there is
no empty spacetime (in the spirit of Mach's principle) and
the spacetime coordinates have a physical meaning only if
there are particles for which the coordinates
can be measured.

Summarizing our discussion, we assume that,
{\it by definition}, on quantum level a Lie algebra is
the symmetry algebra if there exist physical
observables such that their operators
satisfy the commutation relations characterizing the
algebra. Then, a particle is called elementary if the
set of its wave functions is a space of IR of this 
algebra by Hermitian operators.
Such an approach is in the spirit of that considered
by Dirac in Ref. \cite{Dir}. By using the abbreviation
'IR' we will always mean an irreducible representation
by Hermitian operators.    

\section{Galois fields vs. 'infinite' mathematics}
\label{S2} 

The standard mathematics used in quantum physics is essentially
based on the notion of infinity. Although any realistic calculation
involves only a finite number of numbers, one usually believes that
{\it in principle} any calculation can be performed with arbitrary 
large numbers and with any desired accuracy. 

Suppose we wish to verify that 100+200=200+100. In the
spirit of quantum theory it is insufficient to just say
that 100+200=300 and 200+100=300. We should describe an 
experiment where these relations will be verified. In particular,
we should specify whether we have enough resources to
represent the numbers 100, 200 and 300. 

We believe the following observation is very important: 
although standard
mathematics is a part of our everyday life, people typically
do not realize that {\it standard mathematics is implicitly 
based on the assumption that one can have any desirable 
amount of resources}. Also the standard mathematics is based
on some assumptions the validity of which cannot be verified
in principle (e.g. Zorn's lemma or Zermelo's axiom of choice).
This obviously contradicts basic principles of quantum theory. 

\begin{sloppypar}
Suppose, however that our Universe is finite. Then the
amount of resources cannot be infinite and it is 
natural to assume that there exists a
fundamental number $p$ such that all calculations can
be performed only modulo $p$. 
\end{sloppypar}

One might think that division is a natural operation 
since we know from everyday experience that any macroscopic 
object can be divided by two, three and even a million
parts. But is it possible to divide by two or three the electron
or neutrino? It is obvious that if elementary particles exist, then
division has only a limited sense. Indeed, let us consider, for example,
the gram-molecule of water having the mass 18 grams. It contains the
Avogadro number of molecules $6\cdot 10^{23}$. We can divide this 
gram-molecule by ten, million, billion, but when we begin to divide by 
numbers greater than the Avogadro one, the division operation loses its
sense. The obvious conclusion is that the notion of division has a
sense only within some limits. 

What conclusion should be drawn from the above observations? 
We essentially have the following dilemma. The first possibility 
is to accept that standard mathematics is nevertheless suitable 
for describing phenomena with
any numbers but not all of the phenomena can occur in our Universe.
Another possibility is to assume that there exists a number $p$
such that no physical quantity can have a value greater than 
$p$. In that case mathematics describing physics should obviously 
involve only numbers not greater than $p$; in particular 
such a mathematics does not contain the actual infinity at all. 
It is clear that only the second possibility is in the spirit
of quantum theory.

The above dilemma has a well known historical analogy. 
A hundred years ago nobody believed that there exists 
an absolute limit of speed. People did not see any 
reason which in principle does not allow any particle to 
have an arbitrary speed. The special theory
of relativity does not show that the Newtonian mechanics 
is wrong: it is correct but only for phenomena where 
velocities are much less than the velocity of light. 

Our next example is as follows. Suppose we wish to compute
as many decimal digits of $\pi$ as possible and we can 
build a computer as big as the Solar system
which will compute $\pi$ for thousands years. It is clear
that if $N$ is the number of elementary particles in the
Universe we will have no room for writing down $N+1$ decimal 
digits of $\pi$. Therefore all the digits cannot be computed
even in principle.

It is well known that mathematics starts from natural 
numbers (recall the famous Kronecker expression: 
'God made
the natural numbers, all else is the work of man')
which have a clear physical meaning. However only two 
operations are always possible in the set of natural
numbers: addition and multiplication.

In order to
make addition reversible, we introduce negative
integers -1, -2 etc. Then, instead of the set of
natural numbers we get
the ring of integers where three operations
are always possible:
addition, subtraction and multiplication. However,
negative numbers do not have a direct physical
meaning (we cannot say, for example, 'I have
minus two apples'). Their only role is to make
addition reversible.

The next step is the transition to the field of
rational numbers in which all
four operations (except division by zero) are possible.
However, as noted above, division has only a limited
sense.

In mathematics the notion of linear space is
widely used, and such important
notions as the basis and dimension are meaningful only
if the space is considered over a field or body.
Therefore if we start from natural numbers and wish
to have a field, we have to introduce negative
and rational numbers. However, if, instead of all
natural numbers, we consider
only $p$ numbers 0, 1, 2, ... $p-1$ where $p$ is
prime (we treat zero as a natural number) then 
we can easily construct a field without
adding any new elements. This construction, called
Galois field, contains nothing that could prevent
its understanding even by pupils of elementary
schools.

Let us denote the set of numbers 0, 1, 2,...$p-1$
as $GF(p)$. Define addition and multiplication as usual
but take the final result modulo $p$. For simplicity,
let us consider the case $p=5$. Then $F_5$ is the set 0,
1, 2, 3, 4. Then
$1+2=3$ and $1+3=4$ as usual, but $2+3=0$, $3+4=2$ etc.
Analogously, $1\cdot 2=2$,
$2\cdot 2=4$, but $2\cdot 3 =1$, $3\cdot 4=2$ etc.
By definition, the element
$y\in GF(p)$ is called opposite to $x\in GF(p)$ and is
denoted as $-x$ if $x+y=0$ in
$GF_p$. For example, in $GF(5)$ we have -2=3, -4=1 etc.
Analogously $y\in GF(p)$ is
called inverse to $x\in GF(p)$ and is denoted as
$1/x$ if $xy=1$ in $GF(p)$.
For example, in $GF(5)$ we have 1/2=3, 1/4=4 etc. It is
easy to see that
addition is reversible for any natural $p>0$ but for
making multiplication
reversible we should choose $p$ to be a prime.
Otherwise the product of two
nonzero elements may be zero modulo $p$. If $p$ is
chosen to be a prime then
indeed $GF(p)$ becomes a field without introducing any
new objects (like negative numbers or fractions). For
example, in this field each element can obviously be
treated as positive and negative {\bf simultaneously}!

One might say: well, this is beautiful but impractical
since in physics and
everyday life 2+3 is always 5 but not 0. Let us suppose,
however that fundamental
physics is described not by 'usual mathematics' but by
'mathematics modulo $p$'
where $p$ is a very large number.
Then, operating with numbers much smaller than $p$ we
will not notice this $p$,
at least if we only add and multiply. We will notice a
difference between 'usual mathematics' and 'mathematics
modulo p' only while operating with numbers
comparable to $p$.

We can easily extend the correspondence between $GF(p)$ and
the ring of integers $Z$
in such a way that subtraction will also be included.
Since the field $GF(p)$ is cyclic (adding
1 successively, we will
obtain 0 eventually), it is convenient to visually depict
its elements by 
points of a circle of the radius $p/2\pi$ on the plane $(x,y)$.
In Fig. 1 only a
part of the circle near the origin is depicted.
\begin{figure}[!ht]
\centerline{\scalebox{1.1}{\includegraphics{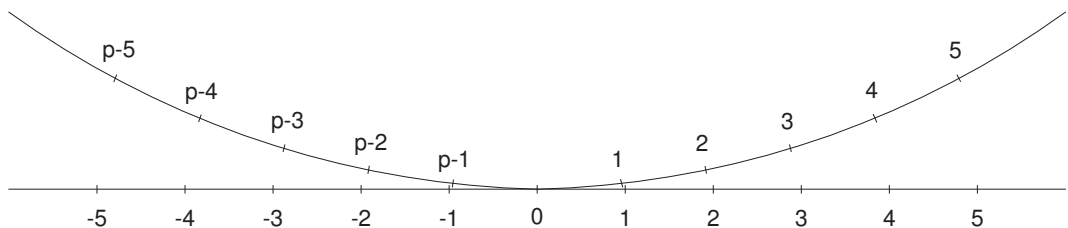}}}%
\caption{%
  Relation between $GF(p)$ and the ring of integers%
}%
\label{Fig.1}
\end{figure}
Then the distance between
neighboring elements of the field is equal to unity, and
the elements
0, 1, 2,... are situated on the circle counterclockwise. At
the same time we depict the elements of $Z$ as usual
such that each element $z\in Z$ is depicted by
a point with the
coordinates $(z,0)$. We can denote the elements of $GF(p)$
not only as 0, 1,... $p-1$
but also as 0, $\pm 1$, $\pm 2,$,...$\pm (p-1)/2$, and such
a set is called the
set of minimal residues. Let $f$ be a map from $GF(p)$ to Z,
such that the element
$f(a) \in Z$ corresponding to the minimal residue $a$ has
the same notation as
$a$ but is considered as the element of $Z$. Denote
$C(p) = p^{1/(lnp)^{1/2}}$ and
let $U_0$ be the set of elements $a\in GF(p)$ such that
$|f(a)|<C(p)$. Then if
$a_1,a_2,...a_n\in U_0$ and $n_1,n_2$ are such natural
numbers that
\begin{equation}
n_1<(p-1)/2C(p),\,\,n_2<ln((p-1)/2)/(lnp)^{1/2}
\end{equation}
\label{0}
then
$$f(a_1\pm a_2\pm...a_n)=f(a_1)\pm f(a_2)\pm ...f(a_n)$$
if $n\leq n_1$ and
$$f(a_1 a_2...a_n)=f(a_1)f(a_2) ...f(a_n)$$ if $n\leq n_2$.

The meaning of the above relations is simple:
although $f$ is not a homomorphism of rings $GF(p)$ and $Z$,
but if $p$ is sufficiently large, then for a sufficiently
large number of elements of $U_0$ the
addition, subtraction and multiplication are
performed according to the same rules
as for elements $z\in Z$ such that $|z|<C(p)$.
Therefore $f$ can be treated as a
local isomorphism of rings $GF(p)$ and $Z$.

The above discussion has a well known historical
analogy. For many years people believed
that our Earth was flat and infinite, and only
after a long period of time they realized that
it was finite and had a curvature. It is difficult
to notice the curvature when we deal only with
distances much less than the radius of the
curvature $R$. Analogously one might think that
the set of
numbers describing physics has a curvature defined
by a very large number $p$ but we do not notice
it when
we deal only with numbers much less than $p$.

Let us note that even for elements from $U_0$ the
result of division in the field
$GF(p)$ differs generally speaking, from the corresponding
result in the field of rational number $Q$. For example the
element 1/2 in $GF(p)$ is a very large number
$(p+1)/2$. For this reason one might think that physics
based on Galois
fields has nothing to with the reality. We will see in
the subsequent section that this is not so since the
spaces describing quantum systems are projective.

Since the standard quantum theory is based on complex
numbers, the question arises whether it is possible
to construct a finite analog of such numbers.
By analogy with the field of complex numbers, we can
consider a set $GF(p^2)$
of $p^2$ elements $a+bi$ where $a,b\in GF(p)$ and $i$ is
a formal element such
that $i^2=1$. The question arises whether $GF(p^2)$ is a
field, i.e.
we can define all the four operations excepting
division by zero.
The definition of addition, subtraction and multiplication
in $GF(p^2)$
is obvious and, by analogy with the field of complex
numbers, one could define division as
$1/(a+bi)\,=a/(a^2+b^2)\,-ib/(a^2+b^2)$ if 
$a$ and $b$ are not equal to zero simultaneously.
This definition can be meaningful only if
$a^2+b^2\neq 0$ in $GF(p)$
for any $a,b\in GF(p)$ i.e. $a^2+b^2$ is not divisible by $p$.
Therefore the definition is meaningful only if
$p$ {\it cannot}
be represented as a sum of two squares and is
meaningless otherwise.
We will not consider the case $p=2$ and therefore $p$
is necessarily odd.
Then we have two possibilities: the value of $p\,(mod \,4)$ 
is either 1 or 3. The well known result of number theory
(see e.g. the
textbooks \cite{VDW}) is that a prime number $p$ can be
represented as a sum of two squares only in the
former case
and cannot in the latter one. Therefore the above
construction of
the field $GF(p^2)$ is correct if
$p=3\,(mod \,4)$.
In that case the above local homomorphism of the rings
$Z$ and $GF(p)$ can be extended to the homomorphism
between the rings $Z+iZ$ and $GF(p^2)$ if we consider
a set $U$ such that $a+bi\in U$ if $a\in U_0$ and 
$b\in U_0$.

The first impression is that if Galois fields can somehow
replace the conventional field of complex numbers then this
can be done only for $p$ satisfying $p=3\,(mod \,4)$ and
therefore the case $p=1\,(mod \,4)$ is of no interest for
this purpose. It can be shown however, \cite{lev2} 
that correspondence between complex numbers and Galois fields
containing $p^2$ elements can also be established if
$p=1\,(mod \,4)$. Nevertheless, arguments given in
Refs. \cite{complex,hep} indicate that if quantum theory 
is based on a Galois
field then $p$ is probably such that $p=3\,(mod \,4)$
rather than $p=1\,(mod \,4)$. In general, it is well known
(see e.g. Ref. \cite{VDW}) that any Galois field consists of
$p^n$ elements where $p$ is prime and $n>0$ is natural.
The numbers
$p$ and $n$ define the field $F_{p^n}$ uniquely up to
isomorphism and $p$ is called the characteristic of the
Galois field.

As discussed in Sect. \ref{S1}, one of the main problems
in modern theory is the existence of infinities. 
A desire to have a theory without divergencies is probably the
main motivation for developing modern theories
extending LQFT, e.g. loop quantum gravity, noncommutative 
quantum theory, string theory etc. For example, the main
idea of the string theory is that a string is a less
singular object than a point and this gives hope that
such a theory will have no divergencies. At the same
time, in GFQT divergencies cannot exist in principle since 
any Galois field is finite.

The idea to replace the field of complex numbers in
quantum theory by something else is well known. There
exists a wide literature where quantum theory is based
on quaternions, p-adic numbers or other constructions.
However, as noted above, if we accept that the future
quantum physics should not contain the actual infinity 
at all then the only possible choice is a Galois field
or even a Galois ring.

\section{Correspondence between standard theory and GFQT}
\label{S3}

The usual requirement for any new theory is that 
at some conditions the theory should reproduce well
known results of the standard theory. In other words,
there should exist a correspondence principle between the
new and standard theories. The existence of such a
principle does not mean of course that the theories
should be absolutely identical; for some phenomena
the predictions of the theories may be essentially
different. The above discussion gives ground to believe
that the standard quantum theory could be treated as a
limit of GFQT when $p\rightarrow \infty$ in the same sense
as classical nonrelativistic mechanics is 
a limit of classical relativistic mechanics when 
$c\rightarrow \infty$ and a limit of nonrelativistic quantum
mechanics when $\hbar\rightarrow 0$. The correspondence
between GFQT and the standard quantum theory has been
discussed in detail in Refs. \cite{lev2,hep} and below
we describe the main ideas of this correspondence. 

A well known historical fact is that originally
quantum theory has been proposed in two formalisms
which seemed to be essentially different: the
Schroedinger wave formalism and the Heisenberg
operator (matrix) formalism. It has been shown later
by Born, von Neumann and others that the both
formalisms are equivalent and, in addition, the
path integral formalism has been developed. A direct
correspondence between GFQT and the standard theory
is rather straightforward if the standard theory is
considered in the operator formalism.    

We define GFQT as a theory where {\it quantum states 
are represented by elements of a 
linear projective space over a field $GF({p^2})$ 
and physical quantities are represented by 
linear operators in that space.} Then a Lie
algebra ${\cal A}$ over $GF(p)$ is called the
symmetry algebra if the operators in $GF(p^2)$
representing the observables belong to a
representation of ${\cal A}$ in $GF(p^2)$.
If this representation is irreducible then
the system is called elementary particle.

At the same time, in the standard theory quantum systems
are described by representations of real Lie algebra in
projective Hilbert spaces. We first have to understand 
how the correspondence between projective Hilbert spaces
and projective linear spaces over $GF(p^2)$ can be
established.
   
The first observation is that Hilbert spaces in quantum
physics contain a big redundancy of elements. Indeed, 
with any desired accuracy any element of the Hilbert 
space can be approximated by a finite linear 
combination 
\begin{equation}
\psi = {\tilde c}_1{\tilde e}_1 + {\tilde c}_2
{\tilde e}_2 + ... {\tilde c}_N{\tilde e}_N
\label{A}
\end{equation} 
where the ${\tilde e}_1,{\tilde e}_2,...$ are the basis 
elements and the 
coefficients ${\tilde c}_1,{\tilde c}_2,...$ 
are complex rational numbers 
(it is well known 
that in any separable Hilbert space the elements (\ref{A}) are 
dense in this space). However, even the set (\ref{A}) contains 
too many elements which are not needed. Indeed, the Hilbert 
spaces in quantum theory are projective, 
i.e. $\psi$ and $c\psi$ represent the same state. 
This is a consequence of the fact that only ratios of 
probabilities are meaningful 
while the probability by itself have no significance. 
In particular, the usual normalization by one is only a 
matter of convenience for 
those who like such a normalization. Therefore we can 
multiply the both parts of Eq. (\ref{A}) by the common 
denominator of the coefficients. In other words, it is 
sufficient to consider only such elements where the 
coefficients have the form ${\tilde c}_j = 
{\tilde a}_j + i{\tilde b}_j$, 
${\tilde a}_j$ and ${\tilde b}_j$ are integers and 
$i$ is the imaginary unity. 

Consider now the elements in $GF(p^2)$, which have the
form
\begin{equation}
x = c_1e_1 + c_2e_2 + ... c_Ne_N
\label{B}
\end{equation} 
such that $f(c_j)={\tilde c}_j$ where the map $f$ is defined
in the preceding section. We also can supply the space over 
$GF(p^2)$ by a scalar product $(x,y)\in GF(p^2)$ such that 
\begin{equation}
(x,y) =\overline{(y,x)},\quad (cx,y)=\bar{c}(x,y),\quad
(x,cy)=c(x,y)
\label{C}
\end{equation}
where the complex conjugation in $GF(p^2)$ is fully
analogous to the standard complex conjugation if
$p=3\, (mod\, 4)$. 
If $f((e_j,e_k))=({\tilde e}_j,{\tilde e}_k)$
then there exists the correspondence between the
elements given by Eqs. (\ref{A}) and (\ref{B}).

Analogously we can define the correspondence between
the operators in projective Hilbert spaces and projective
spaces over $GF(p^2)$ \cite{lev2,hep}. The idea of the
correspondence is rather transparent: we first transform 
the wave functions to make the coefficients in Eq. (\ref{A})
complex integers and if the magnitude of the coefficients
is much less than $p$ than such states are practically
indistinguishable from elements from a linear space
over $GF(p^2)$. If $p$ is very large then there exist 
many states corresponding to each other.

We believe that the above construction also sheds light
on the fact that the notion of probability is a good
illustration of the Kronecker expression. Indeed, 
suppose that we have conducted an experiment $N$ times, 
the first event occurred $n_1$ times, the second - 
$n_2$ times etc. such that $n_1 + n_2 + ... = N$. 
Therefore the experiment is fully described by a set of 
natural numbers. However, people define 
rational numbers $w_i(N) = n_i/N$ and then define the 
limit when $N \rightarrow\infty$.

The idea to apply Galois fields to quantum physics has
been considered by several authors (see e.g. Refs.
\cite{galois}). We believe that our proposal is extremely
natural and straightforward: to take the standard
Heisenberg operator approach to quantum theory and
replace complex numbers by a Galois field. To the best of 
our knowledge, such an approach has not been discussed
in the literature.

\section{Poincare invariance vs. de Sitter invariance}
\label{S4}

The next problem in constructing GFQT is the choice of
the symmetry algebra. Consider first the choice of such an
algebra in the standard theory.

As follows from our definition of symmetry on quantum
level, the standard theory is Poincare invariant if the
representation operators for the system under
consideration satisfy the well-known commutation relations
$$[P^{\mu},P^{\nu}]=0, \quad [M^{\mu\nu}, P^{\rho}]=
-2i(g^{\mu\rho}P^{\nu}-g^{\nu\rho}P^{\mu}),$$
\begin{equation}
[M^{\mu\nu},M^{\rho\sigma}]=-2i (g^{\mu\rho}M^{\nu\sigma}+
g^{\nu \sigma}M^{\mu\rho}-g^{\mu\sigma}M^{\nu\rho}-g^{\nu\rho}
M^{\mu\sigma})
\label{2}
\end{equation}
where $\mu,\nu,\rho,\sigma=0,1,2,3$, $P^{\mu}$ are the
four-momentum operators, $M^{\mu\nu}$ are the representation
operators of the Lorenz algebra and the metric tensor has the nonzero
components $g^{00}=-g^{11}=-g^{22}=-g^{33}=1$.
 
Eq. (\ref{2}) is written in units $\hbar/2=c=1$. Then the spin 
of any particle is integer, the spin of
fermions is odd and the spin of bosons is even. Such a choice
is convenient for establishing correspondence with GFQT (in
units where $\hbar=1$ the spin of fermions is half-integer and,
as noted in Sect. \ref{S3}, the value of 1/2 in a Galois field
is a big number $(p+1)/2$). 

The question arises whether Poincare invariant quantum
theory can be a starting point for its generalization
to GFQT. The answer is probably 'no'
and the reason is the following. GFQT is discrete
and finite because the only numbers it can contain are
elements of a Galois field. Those elements cannot have a dimension 
and operators in GFQT cannot have the continuous spectrum. 

One might argue that quantum physics
should describe the results of measurements and, by definition, any
measurement is performed by a classical observer. So any quantum theory
should necessarily contain three quantities with the dimensions of
mass, time and length.

We believe that a possible counterargument is as follows. Any quantum
theory should contain two parts: 1) a parts describing universal
relations (which do not depend on whether some phenomenon takes place
on the Earth now or at the very early stage of the Universe when
classical measurements were not possible at all); 2) a part describing
how classical measurements are interpreted in terms of those relations.
It is natural to believe that part 1) {\it should not} depend on any dimensional
constants (e.g. $\hbar$, $c$, $G$ etc.) and in particular on whether those
constants are really constant in time. 

Consider, for example such a physical quantity as angular momentum. In units
$\hbar/2=1$ any angular momentum can be only an integer. So we can describe
the value of the angular momentum either by an integer or in units 
$kg\cdot m^2/s$. It is natural to believe that only the first description
is universal while the second one reflects only our macroscopic experience.

The angular momentum operators and Lorenz
boost operators are dimensionless in units $\hbar/2=c=1$ but then the momentum
operators have the dimension of the inverse length.
In addition, the momentum operators and the operators of
the Lorenz boosts contain the continuous spectrum.

Let us recall however the well-known fact that conventional
Poincare invariant theory is a special case of 
de Sitter invariant one. The symmetry algebra of the de Sitter
invariant quantum theory can be either so(2,3) or so(1,4).
Those algebras are the Lie algebras of
symmetry groups of the four-dimensional manifolds in the
five-dimensional space, defined, respectively, as
\begin{equation}
\pm x_5^2+x_0^2-x_1^2-x_2^2-x_3^2=\pm R^2
\label{3}
\end{equation}
where a constant $R$ has the dimension of length.
We use $x_0$ to denote the conventional time coordinate and
$x_5$ to denote the fifth coordinate. The notation $x_5$
rather than $x_4$ is used since in the literature the
latter is sometimes used to denote $ix_0$.

The quantity $R^2$ in the two cases of Eq. \ref{3} 
is often written, respectively, as $R^2=\mp 3/\Lambda$
where $\Lambda$ is the cosmological constant. The
existing astronomical data show that it is very
small. In the literature the latter case is often called the
de Sitter (dS) space while the former is called the
anti de Sitter (AdS) one. 

The both de Sitter algebras are ten-parametric,
as well as the Poincare algebra. However,
in contrast to the Poincare algebra, all
the representation
operators of the de Sitter algebras are dimensionless
(in units $\hbar/2=c=1$). The commutation relations can now be
written in the form of one tensor equation
\begin{equation}
[M^{ab},M^{cd}]=-2i (g^{ac}M^{bd}+g^{bd}M^{cd}-
g^{ad}M^{bc}-g^{bc}M^{ad})
\label{5}
\end{equation}
where $a,b,c,d$ take the values 0,1,2,3,5 and the operators
$M^{ab}$ are antisymmetric. The diagonal metric tensor
has the
components $g^{00}=-g^{11}=-g^{22}=-g^{33}=1$ as usual,
while $g^{55} =1$ for the algebra so(2,3) and
$g^{55}=-1$ for the algebra so(1,4).

When $R$ is very large, the transition from the de Sitter
symmetry to Poincare one (this procedure is called
contraction \cite{IW}) is performed as follows. We define
the operators
$P^{\mu} = M^{\mu 5}/2R$. Then, when
$M^{\mu 5}\rightarrow \infty$, $R\rightarrow \infty$, but
their ratio is finite,
Eq. (\ref{5}) splits into the set of expressions given by
Eq. (\ref{2}).

Note that our definition of the de Sitter symmetry on
quantum level does not involve the cosmological
constant at all. It appears only if
one is interested in interpreting results in terms of
the de Sitter spacetime or in the Poincare limit.
Since all the
operators $M^{ab}$ are dimensionless in units $\hbar/2=c=1$,
the de Sitter invariant quantum theories can be formulated
only in terms of dimensionless variables.

If one assumes that spacetime is fundamental then in the
spirit of GR it is natural to think that
the empty space is flat, i.e. that the cosmological
constant is equal to zero. This was the subject of the
well-known dispute between Einstein and de Sitter
described in a wide literature (see e.g. Refs.
\cite{Mach2,Einst-dS} and references therein). In the
LQFT the cosmological constant
is given by a contribution of vacuum diagrams,
and the problem is to explain why it is so small. On the
other hand, if we assume that symmetry on quantum level in
our formulation is more fundamental, then 
the cosmological constant problem does not arise at all. 
Instead we
have a problem of why nowadays Poincare symmetry is so
good approximate symmetry. 

Summarizing the above discussion, we see that 
elementary particles in GFQT can be investigated by
considering IRs of the so(2,3) or so(1,4)
algebras over a Galois field $GF({p^2})$. The case
so(2,3) has been discussed in detail in Ref. \cite{hep}
and the main results are described below. 

\section{Results and discussion}

The original motivation for investigating GFQT was
as follows. Let us take the standard QED in
dS or AdS space, write the Hamiltonian and other
operators in angular momentum representation and 
replace standard IRs for the electron, positron and
photon by corresponding IRs over $GF({p^2})$. Then we will
have a theory with a natural cutoff $p$ and all
renormalizations will be well defined. In other
words, instead of the standard approach, which,
according to Polchinski's joke \cite{Polchinski}, 
is essentially based on the formula 
'$\infty - \infty = physics$', we
will have a well defined scheme. One might treat 
this motivation as an attempt to substantiate
standard momentum regularizations (e.g. the
Pauli-Villars regularization) at momenta $p/R$
(where $R$ is the radius of the Universe).
In other terms this might be treated as introducing
fundamental length of order $R/p$. We now discuss
reasons explaining why this naive attempt fails.

Consider first
the construction of IR over $GF({p^2})$ for the 
electron. We
start from the state with the minimum energy (where
energy=mass) and gradually construct states with
higher and higher energies. In
such a way we are moving counterclockwise along the 
circle on Fig. 1 in Sect. \ref{S2}. Then sooner or 
later we will arrive at the left half of the
circle, where the energy is negative, and finally 
we will arrive at the point where energy=-mass. 
In other words, instead of the analog of IR describing
only the electron, we obtain an IR describing the 
electron and positron simultaneously. In general the following
conclusion can be drawn: IRs of the 
AdS algebra over a Galois field have the following properties:

i) {\it The representation space of any IR necessarily contains 
states describing both, a particle and its antiparticle. In particular, there are no IRs 
such that their representation space describes only a particle without its 
antiparticle and vice versa;}

ii) {\it There are no IRs describing neutral particles i.e. 
particles which do not have distinct antiparticles.} 

This result is extremely simple and beautiful since it 
shows that {\it in GFQT the very existence of antiparticles
immediately follows from the fact that any Galois field
is finite}. 

In the standard theory a particle and its antiparticle
are described by different IRs but they are combined
together by a local covariant equation (e.g. the Dirac equation). 
We see that in GFQT
the idea of the Dirac equation is implemented without
assuming locality but already at the level of IRs.

Our construction immediately explains why a particle 
and its antiparticle have the same mass and spin but
opposite charges. While in the standard theory this is
a consequence of the CPT theorem, which in particular
involves locality, in GFQT no locality is required. 

One might immediately conclude that since, as a consequence of ii),
the photon in GFQT cannot be elementary, this theory cannot be
realistic and does not deserve attention. 
We believe however, that the nonexistence of 
neutral elementary 
particles in GFQT shows that the photon, the graviton
and other neutral particles should be considered on a
deeper level. For example, several authors considered
a model where the photon is a composite state of Dirac
singletons \cite{FF}. 

In my discussions with
physicists, some of them commented GFQT as 
follows. This is the approach where a cutoff
(the characteristics $p$ of the Galois field) is
introduced from the beginning and for this reason
there is nothing strange in the fact that
the theory does not have infinities. It has a 
large number $p$ instead and this number can be
practically treated as infinite. 

Consider, however the vacuum energy problem. In the standard 
theory the vacuum energy of the electron-positron
field equals $-\infty$. To avoid such an undesirable behavior
it is additionally required that all operators in question
should be taken in normal ordering. However, the requirement
of normal ordering does not follow from the theory; it is simply
an extra requirement aiming to obtain the correct value of the 
vacuum energy. Therefore, if GFQT were simply a theory with a cutoff $p$, 
one would expect the vacuum energy to be of order $p$. However, since the rules
of arithmetic in Galois fields are different, one can prove that \cite{hep}

iii) {\it The vacuum is the eigenvector of all the representation operators with
the eigenvalues zero without imposing an artificial requirement that the 
operators should be written in the normal form.}

This calculation can be treated as the first example when the quantity, 
which in the standard
theory is infinite, is calculated beyond perturbation theory.
The vacuum energy problem is discussed in practically every
textbook on LQFT and it is well known that the result
$E_{vac}=-\infty$ was a motivation for Dirac's hole theory.

The result of GFQT related to the spin-statistics theorem can be formulated 
as follows \cite{hep}

iv) {\it The normal spin-statistics connection simply follows from the
requirement that quantum theory should be based on complex numbers. This is a 
consequence of the famous and elegant fact of number theory that if $p > 2$ 
then -1 can be a square modulo p if and only if p = 1 (mod 4). 
Therefore if p = 3 (mod 4) then the relation $zz* = -1$ can be valid only if 
the residue field $Z/pZ$ is extended.}

Recall that in the standard theory the proof of the theorem involves locality. 
Moreover, there are reasons to believe that GFQT indicates
to a stronger requirement than the spin-statistics theorem:

v) {\it If the numbers of the physical and nonphysical states should 
be the same (in the spirit of our understanding of 
antiparticles) then only fermions can be elementary.}

Such a possibility has been discussed in a wide literature.
In particular, Heisenberg discussed a possibility that
there exists only one fundamental fermion field with the
spin 1/2. In our recent Ref. \cite{jpa} another arguments
are given that only fermions could be elementary.

The results i) - v) are based on the consideration of IRs
of the so(2,3) algebra over a Galois field $GF({p^2})$.
It is also very interesting to investigate IRs of the
so(1,4) algebra over $GF({p^2})$. In particular, a problem
arises whether gravity might be a manifestation of the number $p$.
This problem will be discussed elsewhere. 

We see that GFQT sheds a new light on the fundamental
problems of physics. We believe, however, that not only
this makes GFQT an extremely interesting theory.  
For centuries, scientists and philosophers have been trying to
understand why mathematics is so successful in explaining
physical phenomena (see e.g. Ref. \cite{Wigner1}). 
However, such a branch of mathematics as number 
theory and, in particular, Galois fields, have 
practically no implications in particle physics.
Historically, every new physical theory usually involved
more complicated mathematics. The standard mathematical
tools in modern quantum theory are differential and
integral equations, distributions, analytical functions,
representations of Lie algebras in Hilbert spaces etc.
At the same time, very impressive results of 
number theory about properties of natural numbers 
(e.g. the Wilson theorem) and even the notion of primes 
are not used at all! The reader can easily notice that GFQT 
involves only arithmetic of Galois fields (which are even 
simpler than the set of natural numbers). The very 
possibility that the future quantum theory could be 
formulated in such a way, is of indubitable interest. 
 
\begin{sloppypar}
{\it Acknowledgements:} I am grateful to F. Coester, 
I.E. Dzyaloshinskii, M. Planat, W. Polyzou and M. Saniga for 
discussions and 
to M. Planat for inviting me to this conference.
\end{sloppypar}

\end{document}